\documentclass[aip,twocolumn,amsmath,amssymb,seceqn,10pt]{revtex4-1}
\usepackage{graphicx}% Include figure files
\usepackage{graphics}
\usepackage{epsfig}
\usepackage{dcolumn}% Align table columns on decimal point
\usepackage{bm}% bold math
\usepackage{makeidx}
\usepackage{subfigure}
\usepackage{color}
\usepackage{longtable}

\begin{document}

\title{Focused issue on antiferromagnetic spintronics: An overview \\ (Part of a collection of reviews on antiferromagnetic spintronics)}% Force line breaks with \\
\author{T. Jungwirth}
\affiliation{Institute of Physics, Academy of Sciences of the Czech Republic, Cukrovarnick\'a 10, 162 00 Praha 6, Czech Republic}
\affiliation{School of Physics and Astronomy, University of Nottingham, Nottingham NG7 2RD, United Kingdom}
\author{J. Sinova}
\affiliation{Institut f\"ur Physik, Johannes Gutenberg Universit\"at Mainz, 55128 Mainz, Germany}
\affiliation{Institute of Physics, Academy of Sciences of the Czech Republic, Cukrovarnick\'a 10, 162 00 Praha 6, Czech Republic}
\author{A. Manchon}
\affiliation{Physical Science and Engineering Division, King Abdullah University of Science and Technology (KAUST), Thuwal 23955-6900, Kingdom of Saudi Arabia}
\author{X.~Marti}
\affiliation{Institute of Physics, Academy of Sciences of the Czech Republic, Cukrovarnick\'a 10, 162 00 Praha 6, Czech Republic}
\affiliation{IGS Research Ltd., Calle La Coma, Nave 8, 43140 La Pobla de Mafumet, Tarragona, Spain}
\author{J. Wunderlich}
\affiliation{Hitachi Cambridge Laboratory, Cambridge CB3 0HE,  United Kingdom}
\affiliation{Institute of Physics, Academy of Sciences of the Czech Republic, Cukrovarnick\'a 10, 162 00 Praha 6, Czech Republic}
\author{C. Felser}
\affiliation{Max Planck Institute for Chemical Physics of Solids, N\"{o}thnitzer Str. 40, 01187 Dresden, Germany}

%\date{\today}% It is always \today, today,
             %  but any date may be explicitly specified

%\begin{abstract}

%\end{abstract}

%\keywords{Suggested keywords}%Use showkeys class option if keyword
                              %display desired
\maketitle
%\tableofcontents

{\bf This focused issue attempts to provide a comprehensive introduction into the field of antiferromagnetic spintronics. Apart from the brief overview below, it features five review articles. The intention is to cover in a coherent and complementary way key physical aspects of  the antiferromagnetic spintronics research. These range from microelectronic memory devices and optical manipulation and detection of antiferromagnetic spins, to the fundamentals of antiferromagnetic dynamics in uniform or spin-textured systems, and to the interplay of antiferromagnetic spintronics with topological phenomena. The antiferromagnetic ordering can take a number of forms including fully compensated collinear, non-collinear, and non-coplanar magnetic lattices, compensated and uncompensated ferrimagnets, or  metamagnetic materials hosting an antiferromagnetic to ferromagnetic phase transition. Apart from the variety of distinct magnetic crystal structures, the focused issue also encompasses spintronic phenomena and devices studied in  antiferromagnet/ferromagnet heterostructures and in synthetic antiferromagnets.}

Ferromagnets have been a subject of fascination and have been of great practical value for thousands of years. They were in the cradle of sound-recording more than a hundred years ago and carried on through the video-recording era to the modern data-storage media and computer memories. The latest incarnation of magnetic recording and magnetic sensors, called spintronics, was also all due to ferromagnets\cite{Wolf2001}.

\begin{figure}[h]
	\centering
	\includegraphics[width=1\columnwidth]{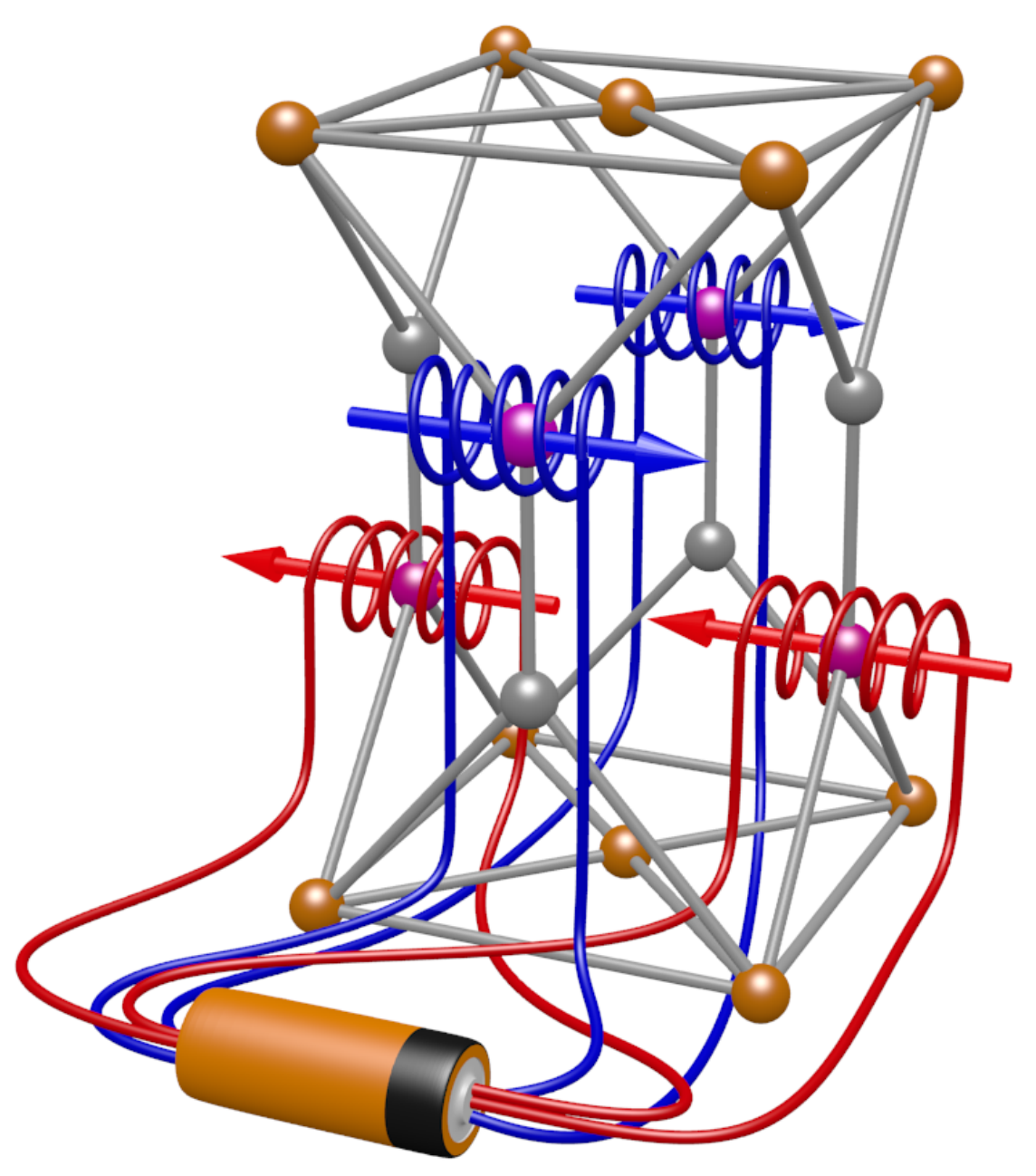}
	\caption[fig_1]{Within traditional schemes of magnetic recording, efficient control of antiferromagnets would require generating local Oersted fields of alternating sign, commensurate with the staggered N\'eel order. While winding a micro-coil around each individual magnetic atom is unlikely to become a meaningful concept, antiferromagnetic spintronics is discovering new physical principles bringing this and other seemingly science-fiction ideas into reality.}
	\vspace{-0.2 cm}
	\label{fig_1}
\end{figure}

Antiferromagnets are known from the 1930's as the second basic type of magnetic order\cite{Neel1970}. They are abundant, yet practical means combining manipulation and detection of antiferromagnetic spins in microelectronic devices have remained elusive (see Fig.~1), until a year ago\cite{Marrows2016}. Among the key prerequisites that led to the 2016 experimental demonstration of the antiferromagnetic memory device were studies of relativistic charge-spin coupling phenomena that have shaped much of the recent research and development in spintronics\cite{Pulizzi2012}. These relativistic spintronic phenomena were initially studied in non-magnetic systems but soon after led to discoveries of new highly efficient means for controlling ordered spins in ferromagnets. However, since these mechanisms are built on physical principles that are independent of the magnetic order, they have opened the door for exploiting  also antiferromagnets in microelectronics. 

Over the past year, the memory functionality has been verified in epilayers of semimetallic antiferromagnetic compounds deposited on silicon and other common semiconductor wafers, as well as in sputtered metallic antiferromagnetic films.  The microfabricated antiferromagnetic bit-cells are compatible with standard microelectronic circuitry. Multi-level (neuron-like) switching characteristics that allow to integrate memory and logic within the bit-cell have been observed in antiferromagnetic metals, semimetals, and semiconductors. Contact as well as non-contact electrical writing has been demonstrated with writing pulse-length downscaled from seconds to picoseconds.  Antiferromagnetic memories, together with a broader overview of electrical methods for manipulating and detecting  antiferromagnetic spins, are discussed in the article by Zelezny {\em et al.}\cite{Zelezny2017x}.

Initial studies of the optical control of spins in antiferromagnets preceded by a decade the realization of antiferromagnetic memories in microelectronic devices.  However, the research in antiferromagnetic opto-spintronics has also significantly intensified recently and the range of studied materials has extended beyond insulating antiferromagnets to include conductive materials. Apart from the growing materials overlap, both optical and electrical control of antiferromagnetic spins can now explore and utilize the ultra-fast THz-scale spin dynamics in antiferromagnets. Detection tools used in optics and electronics are also closely  linked. For example, the frequently employed magneto-optical effects are ac counterparts of the dc magneto-transport phenomena used for the readout in electronic antiferromagnetic devices. However, many aspects of the opto-spintronics research are unique to the control of antiferromagnetic moments by light.  The article by Nemec {\em et al.}\cite{Nemec2017x} reviews all these key aspects of the antiferromagnetic opto-spintronics, starting from the optical detection methods and continuing to discuss the THz emission, ultra-fast optical switching, and the time-resolved spin-dynamics measurements. 

Recent breakthroughs in the efficient manipulation and detection of antiferromagnetic spins unlock a multitude of well known and newly emerging unparalleled features of antiferromagnets. The THz-scale spin dynamics is an example of an outstanding property of antiferromagnets acknowledged already since the 1950's seminal work by Kittel \cite{Kittel1951}. This can now be exploited in the development of magnetic memories for the ultimate THz band.  Gomonay {\em et al.}\cite{Gomonay2017x} provide in their commentary an overview of the phenomenology of the switching and spin-resonance in antiferromagnets induced by the current-induced spin-torques. The article discusses uniform antiferromagnets as well as antiferromagnetic nanostructures and spin-textures. It also reviews recent studies  of a highly efficient spin-transport driven by antiferromagnetic spin-waves. 

A counter-example of a research avenue that could not have been envisaged at the time of N\'{e}el's or Kittel's seminal works is the topological antiferromagnetic spintronics. Topologically protected Dirac, Weyl, or Majorana quasiparticles and their interplay with the spintronic control of antiferromagnetic moments are concepts that have been proposed only very recently and still await experimental verification. Their potential utility ranges from realizing highly efficient magnetoresitive readout and spin-torque writing schemes in antiferromagnetic memories to robust topological qubit. This emerging research avenue is reviewed by Smejkal {\em et al.}\cite{Smejkal2017x}. The article pays a special attention to spin-transport phenomena characterized by topological invariants in the momentum or real space.  Collinear, non-collinear, and non-coplanar antiferromagnets and textures are considered to cover the broad range of topological effects that may occur in antiferromagnets. 

Throughout the articles of this focused issue, antiferromagnetic and ferromagnetic spintronics fields are linked to highlight their similarities and, more importantly, their distinct features. 
The inherent radiation hardness and non-volatile memory in the ordered spin state are assets common to both antiferromagnetic and ferromagnetic devices. Similarly, effects that are even in the (sublattice) magnetic moment, like the anisotropic magnetoresistance or the optical magnetic linear dichroism, are equally present in the two types of magnetic order. On the other hand, the insensitivity to magnetic fields and no fringing stray fields make antiferromagnets potentially more favorable for highly integrated memories with close-packed bit cells. The THz antiferromagnetic resonance scales shift the threshold speeds of energy efficient spin manipulation  by orders of magnitude compared to ferromagnets. The effective time-reversal symmetry, in which the time-reversal combined with other crystal symmetry like spatial translation or inversion is a symmetry operation of the magnetic lattice, is another fundamental feature making antiferromagnets distinct from ferromagnets. This combined with the possibility of more complex non-collinear or non-coplanar magnetic orders allows for much richer synergies of spintronics and topological effects in antiferromagnets than in ferromagnets.

The links between antiferromagnetic and ferromagnetic spintronics highlighted in this focused issue also illustrate how one type of magnetic order can benefit from the other in spintronic devices. While  Zelezny {\em et al.}\cite{Zelezny2017x} include in their overview memory devices fabricated in antiferromagnet/ferromagnet heterostructures and in metamagnetic materials, and Gomonay {\em et al.}\cite{Gomonay2017x} remark on ferrimagnets, the field of synthetic antiferromagnets is covered in a dedicated article by Duine {\em et al.}\cite{Duine2017x}. These heterostructures comprising multilayers of antiferromagnetically coupled ferromagnetic films are discussed separately because the carrier mediated antiferromagnetic coupling through a non-magnetic spacer is of a physically distinct origin than the inter-atomic exchange coupling in crystal antiferromagnets. Moreover, synthetic antiferromagnets played a decisive role from the early research and development of spintronic sensor and memory devices and their studies got a new impulse from recent discoveries in relativistic spintronics and emergent concepts in topological spintronics. 

The articles in this issue concentrate on fundamental physics aspects of antiferromagnetic spintronics. Despite its infancy it is, however, timely to frame the field also in the context of its potential future applications. As of 2016, the Moore's Law driven International Technology Roadmap for Semiconductors
is officially at an end\cite{Waldrop2016}. It is being replaced with the new International Roadmap for Devices and Systems in order to tackle the semiconductor scaling problem that is further magnified by the huge increase in the complexity of information technologies. It is envisaged that new material types and new device concepts will join forces with semiconductors in the omnipresent data collection, processing and storage devices. This might help in matching their massive deployment with the unavoidable functional diversity and energy efficiency. Non-volatile multilevel neuron-like bit cells,  the ultra-fast operation and robustness against external perturbations, or alternative concepts based on coding information in robust antiferromagnetic textures  are among the examples that might bring the antiferromagnetic order into the spotlight of the frontier research and development in the "Beyond Moore" technology era.

%The authors acknowledge the Alexander von Humboldt Foundation, the ERC Synergy Grant SC2 (No. 610115), the Transregional Collaborative Research Center (SFB/TRR) 173 SPIN+X, the Grant Agency of the Czech Republic Grant No. 14-37427G, and  the Ministry of Education of the Czech Republic Grant No. LM2015087.

%\bibliography{refs,Notes}
%merlin.mbs aipnum4-1.bst 2010-07-25 4.21a (PWD, AO, DPC) hacked
%Control: key (0)
%Control: author (8) initials jnrlst
%Control: editor formatted (1) identically to author
%Control: production of article title (0) allowed
%Control: page (1) range
%Control: year (1) truncated
%Control: production of eprint (0) enabled
%

\end{document}